\documentclass[prb,reprint,aps,twocolumn]{revtex4-1}

\pdfoutput=1

\usepackage[utf8]{inputenc}

\usepackage[T1]{fontenc}
\usepackage{hyperref}
\usepackage{color}
\usepackage{graphicx}
\usepackage{units}
\usepackage{amsmath}
\usepackage{float}

\usepackage{epstopdf}
\usepackage{subcaption}
\usepackage{amssymb}
\usepackage{pifont}
\usepackage{textcomp}
\usepackage[usenames,dvipsnames]{xcolor}
\usepackage{ulem}
\usepackage{amsmath}

\hypersetup{colorlinks = true, citecolor = blue, breaklinks = true}

\begin{document}

\title{Epitaxial Strain-Dependence of Band Gaps in Oxynitrides Compared to Oxides}
\date{\today}
\author{Nathalie Vonr\"uti}
\affiliation{Department of Chemistry and Biochemistry, University of Bern, Freiestrasse 3, CH-3012 Bern, Switzerland}
\author{Ulrich Aschauer}
\affiliation{Department of Chemistry and Biochemistry, University of Bern, Freiestrasse 3, CH-3012 Bern, Switzerland}

\begin{abstract}
Perovskite oxynitrides are a promising class of material for photocatalytic water-splitting due to their small band gaps and suitably aligned band edges. Recently, epitaxially strained oxynitrides started to attract interest due to the possibility to engineer the anion order and ferroelectric distortions. However, contrary to oxides there have been no studies on the band-gap evolution in oxynitrides with epitaxial strain. Here, we investigate, using density functional theory calculations, the influence of epitaxial strain on the band gap of two different oxynitrides and compare our results with oxides. As opposed to cubic oxides, where both compressive and tensile strain narrow the band gap, we find that the anion order leads to non-degenerate N $2p$ bands already at zero strain, which leads to opposite and anion order-dependent evolutions of the band gap under compressive and tensile strain. Further, we find that in oxynitrides polar distortions increase the band gap by up to 1 eV for 4\% strain whereas for oxides the increase is only about 0.1 eV for the same amount of strain. The reason for this difference is the larger ionic radius of nitrogen leading to larger ferroelectric distortions and a stronger dependence of the band gap on the distortion amplitude. These results imply that ferroelectric distortions are strongly detrimental to light absorption and that their suppression, for example above a critical temperature, should result in a marked decrease in band gap and enhanced absorption of visible light.
\end{abstract}

\maketitle

\section{Introduction}

\begin{figure}
	\includegraphics[width=0.9\columnwidth]{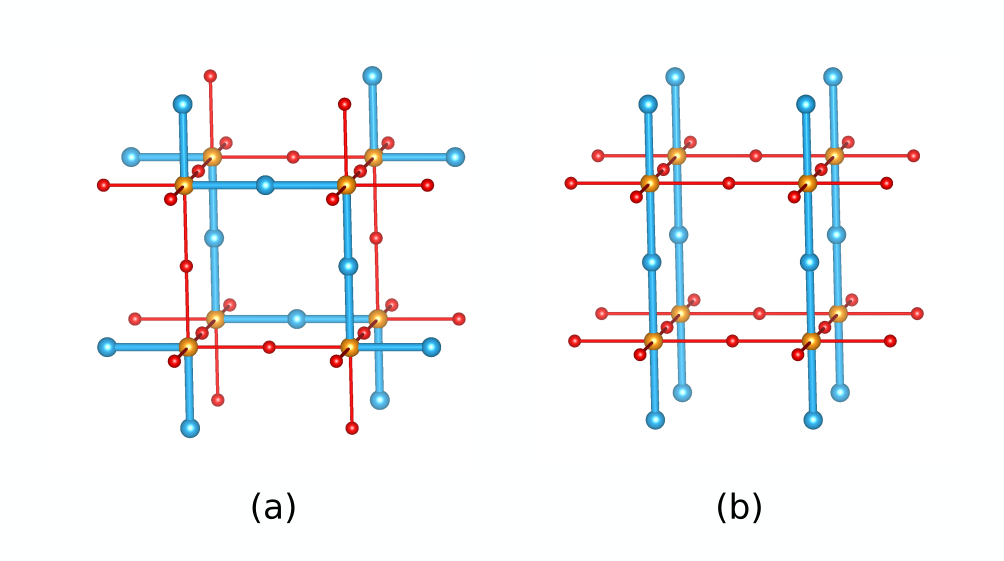}
	\caption{The (a) \textit{cis} and (b) \textit{trans} anion orders of ABO\textsubscript{2}N perovskite oxynitrides considered in this work. Color code: B site=orange, nitrogen=blue, oxygen=red. The A site is not shown.}
	\label{fig:anion_order}
\end{figure}

Perovskite oxynitrides have seen increased interest in recent years due to their absorption in the visible part of the solar spectrum as well as their band edges that straddle the oxygen and hydrogen evolution potentials, making them very promising photoelectrodes for photocatalytic water-splitting \cite{fuertes2012chemistry, Takata2015, Pan2015}. The lower electronegativity of nitrogen compared to oxygen results in N $2p$ states that are higher in energy than the O $2p$ states that form the valence-band edge in oxynitrides and oxides respectively. Compared to pure oxides \cite{fujishima1972electrochemical} that typically have band gaps larger than 3 eV, the partial nitrogen substitution results in reduced band gaps of oxynitrides that are just above 2 eV \cite{Kasahara:2002kd}. A further difference between oxides and oxynitrides are the additional anionic structural degrees of freedom of the latter. The anions in most oxynitrides are only locally ordered, in the bulk preferring a so called \textit{cis} anion order (see Fig. \ref{fig:anion_order} a), which maximizes the overlap between the N \textit{2p} and the transition metal (TM) $d$ orbitals \cite{clarke2002oxynitride, yang2011anion, fuertes2012chemistry, attfield2013principles, ninova2017surface}.

The smaller band gap of oxynitrides compared to oxides is one of the main reasons behind the large interest in these materials for photocatalytic water-splitting. The band gap was shown to depend on the anion order \cite{kubo2017anion, kikuchi2017fundamental, ziani2017photophysical, kubo2017mgtao2n} due to differences in the covalence of the TM-X (X=O,N) $\pi$ p-d interaction. This covalence is stronger in 3D nitrogen orders leading to a stabilization and thus a lowering of the valence band maximum (VBM) for these structures \cite{kubo2017anion}. Octahedral rotations were also shown to affect the band gap via changes of the band dispersions as a function of TM-X-TM bond angles, a cubic structure with 180$^\circ$ bond angles having the smallest band gap \cite{aguiar2008vast, Balaz:2013bpa, Pichler:2017cp}.

Epitaxial strain was shown to alter the anion order in oxynitrides, large compressive strain leading to a \textit{trans} order, where N-TM bonds form 1D chains along the out-of-plane direction (see Fig. \ref{fig:anion_order} b), while large tensile strain favors a \textit{cis} anion order with N-TM bonds along the in-plane directions \cite{oka2014possible, oka2017strain, vonruti2017anion}. Additionally, we observed that epitaxial strain can induce ferroelectric distortions, the direction of which is correlated with the anion order \cite{vonruti2017anion} as both N-TM bonds and ferroelectric distortions preferentially occur along the elongated directions of the epitaxially strained material: Nitrogen ions prefer to orient along elongated directions due to their larger radius compared to oxygen, while ferroelectric distortions typically occur also along the elongated direction due to the reduced Coulomb force that counteracts the ferroelectric distortion \cite{Bersuker:2013fa, Rondinelli:2009tq}.

The effect of epitaxial strain and strain-induced ferroelectricity on the band gap have however only been explained for oxides so far. Berger et al.\cite{berger2011band} analyzed these effects for the case of SrTiO\textsubscript{3} and found that 4\% compressive strain lowers the band gap of the paraelectric phase by approximately 0.1 eV, while the band gap increases by roughly the same amount in the ferroelectric phase. This different behaviour of the paraelectric and the ferroelectric phase stems from two different effects occurring under epitaxial strain as discussed in the following paragraphs.

At zero strain, the paraelectric phase is cubic (in the room-temperature phase without antiferrodistortive octahedral rotations) and the energies of the TM $t_{2g}$ orbitals forming the conduction band are degenerate as are the O $2p$ orbitals forming the valence band. Applying epitaxial strain in the $xy$-plane changes the energy of orbitals containing $x$ and $y$ components relative to orbitals containing z components. Under compressive strain, orbitals without a $z$ component rise in energy, resulting in a narrowed band gap between raised O $2p_{x}$/$p_{y}$ and almost unaffected TM $d_{xz}$/$d_{yz}$ states. Under tensile strain, the orbitals without a $z$ component drop in energy and the gap is between almost unaffected O $2p_{z}$ and lowered TM $d_{xy}$ orbitals. Thus both compressive and tensile strain reduce the band gap in oxides, under compressive strain by raising the valence-band edge, and under tensile strain by lowering the conduction-band edge. 

The ferroelectric distortion on the other hand leads to a stabilization of the O $2p$ orbitals forming the valence-band edge, along with a destabilization of the TM $t_{2g}$ $\pi^*$ orbitals forming the conduction-band edge, resulting in an opening of the band gap \cite{Wheeler:1986dd}. The magnitude of the band gap in a strained oxide material will hence depend on the strain-dependent existence and amplitude of the ferroelectric distortion. The same is true for rotations of the oxygen octahedra, that also lead to an opening of the band gap \cite{berger2011band}.

To assess the prospects of strain-engineered oxynitrides for photocatalysis, we predict in the present work the strain dependence of the band gap in LaTiO\textsubscript{2}N and SrTaO\textsubscript{2}N oxynitrides using density functional theory (DFT) calculations. Since the band gap in oxynitrides additionally depends on the anion order, which is also affected by strain, oxynitrides are expected to exhibit a significantly more complex strain dependence of the band gap than oxides. We hence study the observed changes in band gap starting from calculations in the high symmetry phase and increase the complexity by sequentially adding ferroelectric and octahedral distortions and compare our results for oxynitrides with oxides. Our results show that already at zero strain the anion order leads to non-degenerate N $2p$ bands that form the valence band edge, which leads to opposite evolutions of the band gap under compressive and tensile strain. We also show that strain and anion-order dependent polar distortions lead to a much larger widening of the band gap in oxynitrides than in oxides. 

\section{Computational methods}

Our DFT calculations were performed with the Quantum ESPRESSO package \cite{giannozzi2009quantum} using the gradient corrected PBE exchange-correlation functional \cite{perdew1996generalized} and ultrasoft pseudopotentials \cite{vanderbilt1990soft} with Sr(4s,4p,5s), La(5s,5p,5d,6s), Ti(3s,3p,3d,4s), Ta(5s,5p,5d,6s,6p), O(2s,2p) and N(2s,2p) electrons as valence states. The cutoff for the plane-wave basis set was 40 Ry and 320 Ry for the kinetic energy and the augmented density respectively. Calculations for \textit{trans}-ordered oxynitrides and SrTiO\textsubscript{3} without octahedral rotations were performed using a 5-atom unit cell. All other structures with \textit{cis} order and/or octahedral rotations were computed using a 40-atom unit cell. Reciprocal space integration was performed using Monkhorst-Pack grids \cite{monkhorst1976special} of dimension $8\times 8\times 8$ for the 5 atom unit cells and $4\times 4\times 4$ for the 40 atom $2\times 2\times 2$ supercells.
 
 For LaTiO\textsubscript{2}N we applied a Hubbard U \cite{anisimov1991band} of 3 eV as we did in our previous study \cite{vonruti2017anion}, while no U was applied for SrTiO\textsubscript{3} to have a better comparison with the LDA results of Berger et al. \cite{berger2011band}. We did also not apply a U for SrTaO\textsubscript{2}N as the electronic structure is not strongly affected by changes in U for the more diffuse $5d$ states (see supplementary Fig. S1a). These slight differences in computational setup and the general underestimation of band gaps at the semi-local level of theory will not affect our conclusions, as we are interested in relative geometry-dependent changes of band gaps that do not depend strongly on the level of theory \cite{berger2011band} and U (see supplementary Fig. S1b). 

For geometry relaxations of materials epitaxially strained perpendicular to one of the pseudocubic directions, we assume a cubic substrate: We define the magnitude of strain with respect to the unstrained area in the strain plane of the given anion order and distortion mode(s) and set the in-plane lattice parameters to equal lengths and orthogonal to each other. We then relax the out-of-plane lattice parameter and all internal coordinates until the forces converge below 0.05 eV/\AA\ and total energies change by less than $1.4\cdot 10^{-5}$ eV. Atomic structures were visualized using VESTA \cite{momma2011vesta}.

\section{Results and Discussion}

\subsection{Band gaps in low-energy phases of strained LaTiO\textsubscript{2}N and SrTaO\textsubscript{2}N}
\label{low_energy}

\begin{figure*}
	\centering
	\includegraphics[width=0.95\textwidth]{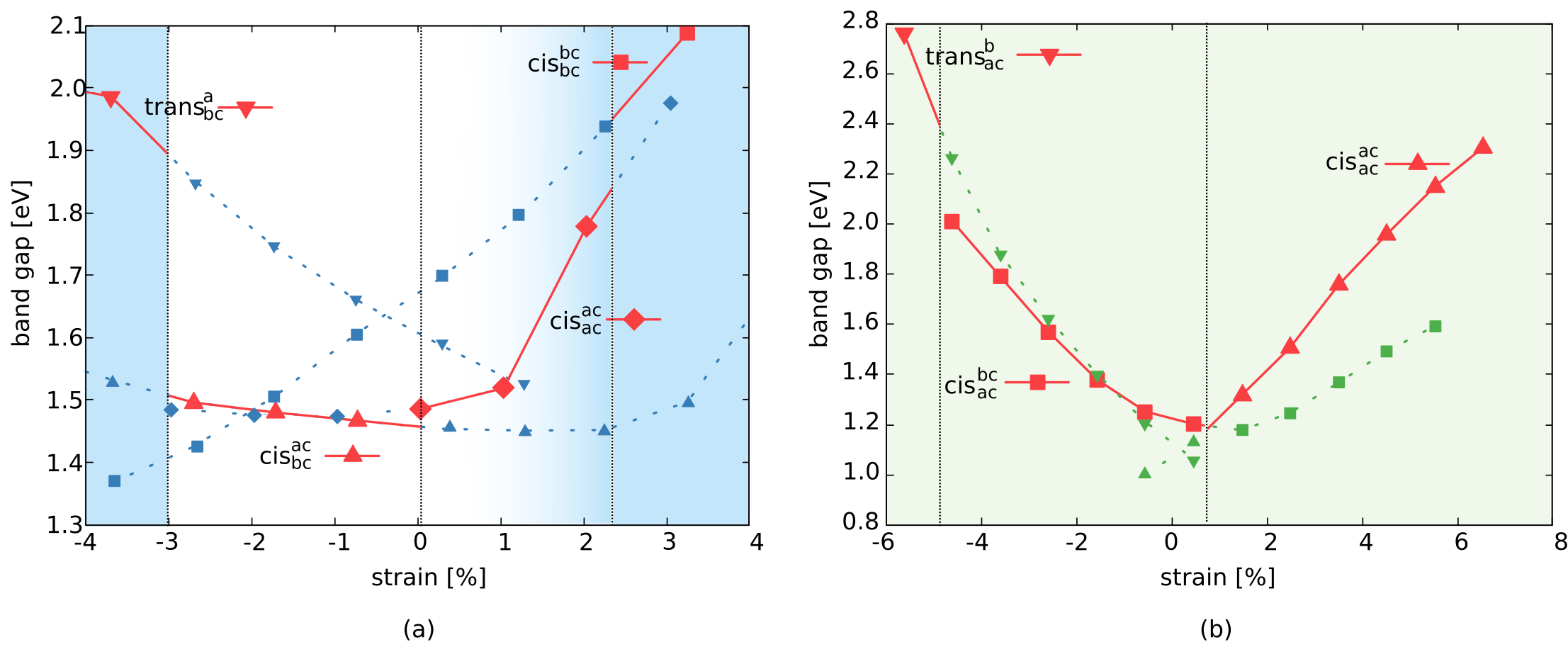}
	\caption{Strain-dependent band gaps for the thermodynamically most stable phases of (a) LaTiO\textsubscript{2}N and (b) SrTaO\textsubscript{2}N. The vertical dashed lines indicate borders between strain ranges for which a certain structure is stable. The band gap of the most stable structure in a specific strain range is colored red. Polarity of the structures is indicated by the background: A white background indicates that the most stable structure is non polar while a colored background indicates that the most stable structure is polar. In this figure a slightly different definition of strain compared to the rest of the article is used: we define strain with respect to the lowest-energy unstrained anion order (i.e. the \textit{cis} structure with a rotation pattern $a^-b^-c^+$ and N-TM bonds along a and b in the case of SrTaO\textsubscript{2}N).}
	\label{fig:bandgap_real}
\end{figure*}

In Fig. \ref{fig:bandgap_real}, we show the epitaxial strain-dependent GGA(+U)-band gaps in fully relaxed LaTiO\textsubscript{2}N and SrTaO\textsubscript{2}N films with different anion orders and strain directions. To differentiate between the combinations of anion orders and strain directions, we use the same notation as in our previous work \cite{vonruti2017anion}: $a_s^b$, where $a$ denotes the anion order (\textit{cis}/\textit{trans}), $b$ the direction/plane along/in which the N-TM-N bonds are aligned and $s$ the strain plane. For example \textit{cis}$^{ac}_{bc}$ refers to a \textit{cis} anion order as shown in Fig. S2 in the supplementary, where N-TM-N chains run in $a$ and $c$ direction with epitaxial strain applied along $b$ and $c$. We will focus here only on anion orders that are energetically most stable somewhere in the considered strain range. For LaTiO\textsubscript{2}N this is discussed in detail in our previous work \cite{vonruti2017anion}, while for  SrTaO\textsubscript{2}N a similar analysis is given in the supplementary material section S2.

In these strained structures, there is a simultaneous strain-dependent evolution of the octahedral rotation angles, polar distortions and the anion order, all of which may have an effect on the band gaps. In order to rationalize the observed band gaps, we will first discuss the strain-dependent structure before we return to the evolution of the band gaps.

Structural features such as the anion order and the existence of a polarity as a function of strain are shown in Fig. \ref{fig:bandgap_real} as labels and background shading respectively. We find that LaTiO\textsubscript{2}N and SrTaO\textsubscript{2}N behave similarly under large compressive and tensile strain, where we observe respectively a \textit{trans} structure with TM-N bonds and ferroelectric distortions oriented out-of-plane and a \textit{cis} order with TM-N bonds and a 2D ferroelectric distortion oriented in-plane  with half of the transition metals being displaced towards the nitrogen in one direction and the other half displaced along the other direction. In the intermediate strain range we see differences in the anion orders and the polarity between the two materials: In SrTaO\textsubscript{2}N, $cis_{ac}^{bc}$, which is thermodynamically stable in an intermediate strain range, shows the same 2D polarity as the structure that is stable in the tensile strain range. This polar distortion pattern is in agreement with previous studies of MgTaO\textsubscript{2}N \cite{kubo2017mgtao2n} and CaTaO\textsubscript{2}N \cite{kubo2017anion}, where asymmetric bond lengths between consecutive Ta-X (X=O, N) bonds (i.e. Ta-X-Ta) were observed in the bulk structures. These structural features can be interpreted as ferroelectric distortions and explained by the partial covalent nature of the Ta-X bond, mainly associated with a d-p $\pi$ interaction \cite{yang2011anion, clark2013thermally, porter2013study, fuertes2015metal}. Recently, also in $cis$ SrTaO\textsubscript{2}N ferroelectricity was measured experimentally in the bulk \cite{kikkawa2016ferroelectric}. In the intermediate-strain range of LaTiO\textsubscript{2}N on the other hand, the low energy structures  ($cis_{bc}^{ac}$ and $cis_{ac}^{ac}$) show no polar distortions. The covalent bond of $5d$ TM-O $2p$ is generally said to be stronger than the $3d$ TM-O $2p$ bond \cite{mingos1998essential}. We therefore argue that the stronger covalent bond formation in SrTaO\textsubscript{2}N favours the ferroelectric distortion more than the weaker covalent bond formation in LaTiO\textsubscript{2}N. 

Despite the similar tolerance factors\cite{goldschmidt1926gesetze} (see supplementary section S3), we observe significantly smaller octahedral rotation angles in SrTaO\textsubscript{2}N compared to LaTiO\textsubscript{2}N (see supplementary Fig. S4). At the same time SrTaO\textsubscript{2}N shows larger polar distortions throughout the whole range of strain (shown by shaded backgrounds in Fig. \ref{fig:bandgap_real} and in supplementary Fig. S5), whereas polar distortions are weaker in LaTiO\textsubscript{2}N and even absent in the structures that are stable at intermediate strain ranges. This inverse correlation of ferroelectricity and octahedral rotations is in agreement with the generally observed competition of these two distortions \cite{Benedek:2013jl, aschauer2014competition}. Displacements of the structures along unstable eigenmodes confirm these observations: while in LaTiO\textsubscript{2}N octahedral rotations lower the energy of the relaxed high-symmetry cell by approximately the same amount (0.045 eV/f.u.) as a ferroelectric distortion (0.048 eV/f.u.), in SrTaO\textsubscript{2}N the energy lowering associated with octahedral rotations (0.028 eV/f.u.) is clearly smaller than that of the ferroelectric distortion (0.090 eV/f.u. eV). We can thus explain the smaller octahedral rotations of SrTaO\textsubscript{2}N by its larger ferroelectric distortions, which compete with the octahedral rotations.

Having discussed the evolution of anion orders and polar distortions as well as differences in octahedral rotations between the two materials, we now return to the band gaps shown in Fig. \ref{fig:bandgap_real}. We want to stress here that since we applied a Hubbard U to LaTiO\textsubscript{2}N but not to SrTaO\textsubscript{2}N, the absolute values of the band gap should not be compared. Reliable values for the absolute band gap require anyway more sophisticated (but also computationally more expensive) methods such as hybrid functionals or quasi-particle GW approaches. Relative changes of the band gaps that are relevant for our study, are however meaningful as validated by these methods \cite{berger2011band}. Overall the evolution of the band gaps is similar in the two materials, as we find larger band gaps with increasing compressive or tensile strain compared to smaller band gaps in the unstrained structures. The magnitude of band-gap changes in SrTaO\textsubscript{2}N is however larger than that in LaTiO\textsubscript{2}N and as opposed to LaTiO\textsubscript{2}N, we do in SrTaO\textsubscript{2}N not see any low energy structures that lead to band gaps which are fairly insensitive to strain. This different behaviour seems to be linked to the polarity of the respective structures. In LaTiO\textsubscript{2}N (Fig. \ref{fig:bandgap_real}a) structures with rapidly changing band gap as a function of strain (\textit{trans}$_{bc}^{a}$ and \textit{cis}$_{bc}^{bc}$) are polar, while those with slowly varying band gaps (\textit{cis}$_{bc}^{ac}$ and \textit{cis}$_{ac}^{ac}$) are non-polar. The kink in the band gap for $cis\mathrm{_{ac}^{ac}}$ between 0\% and 2\% tensile strain is linked to the onset of a 2D ferroelectricity at 1\% strain and further supports this hypothesis. This classification is also in agreement with our results for SrTaO\textsubscript{2}N (see Fig. \ref{fig:bandgap_real} b), where all low energy structures are polar and rapidly change their band gap with strain.

These results highlight two main differences in the strain-dependence of band gaps in oxynitrides compared to oxides. First, the band gap in non-polar oxides decreases in both strain directions \cite{berger2011band}, while in oxynitrides the band gap decreases for some non-polar phases and increases for others (compare \textit{cis}$_{bc}^{ac}$ and \textit{cis}$_{ac}^{ac}$ in LaTiO\textsubscript{2}N). Second, oxynitrides have a much larger increase of the band gap (almost 1 eV for 4\% strain) for polar phases than oxides (0.1 eV for 4\% strain) \cite{berger2011band}. To explain these differences, we will in the following sections investigate the increase in band gaps in oxynitrides for the high symmetry \textit{cis} and \textit{trans} structure before increasing the complexity by first considering polar distortions and then octahedral rotations. With the knowledge gained in those sections we will then conclude on the origin of these differences compared to oxides.

\subsection{High-symmetry phase}\label{para}

\begin{figure}
	\includegraphics[width=0.9\columnwidth]{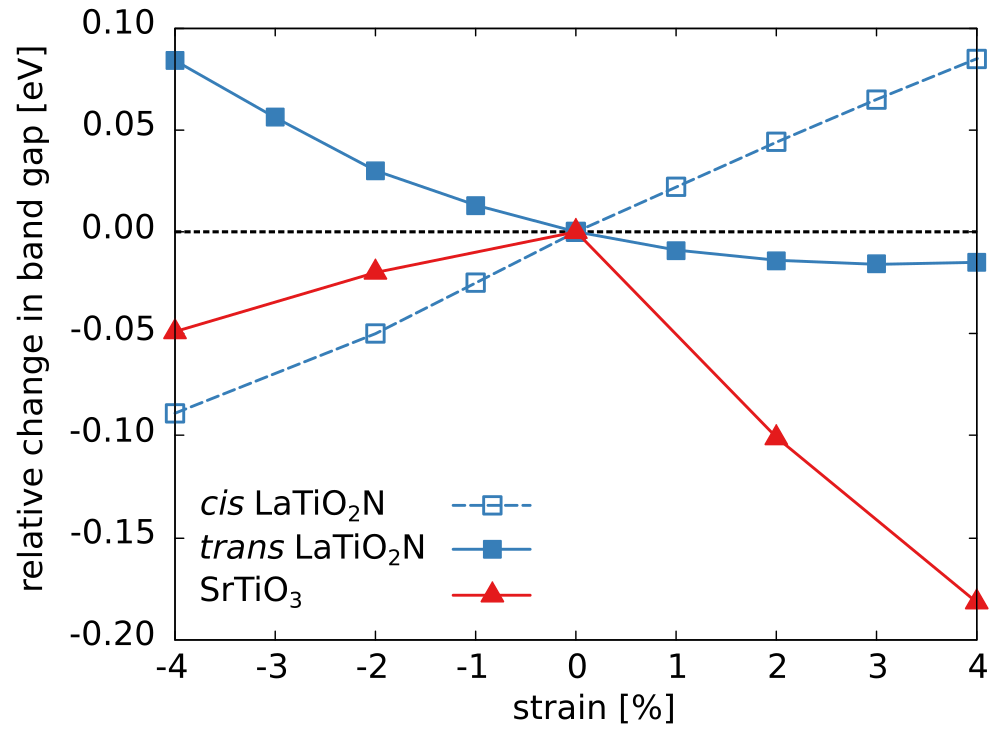}
	\caption{Relative change in band gap for high-symmetry \textit{cis}-LaTiO\textsubscript{2}N, \textit{trans}-LaTiO\textsubscript{2}N and SrTiO\textsubscript{3}, the absolute GGA(+U) band gaps at zero strain are 0.2 eV, 0.6 eV and 1.9 eV respectively.}
	\label{fig:bandgap}
\end{figure}

\begin{figure}
	\includegraphics[width=\columnwidth]{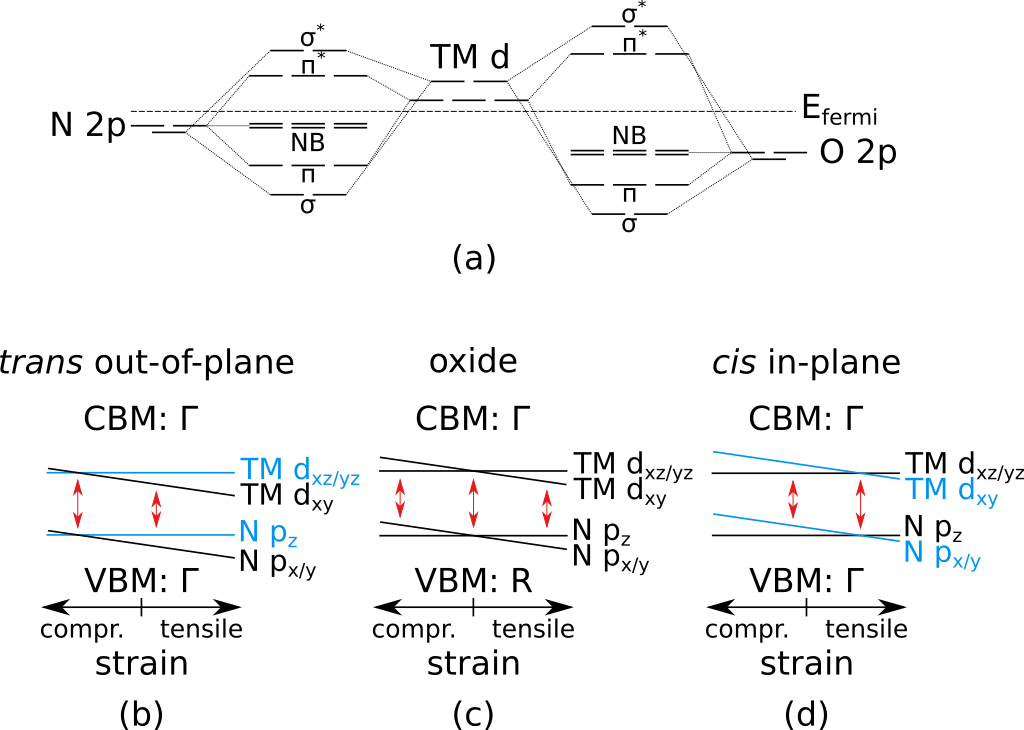}
	\caption{(a) Schematic molecular orbital diagram of a MN\textsubscript{6} (left) and MO\textsubscript{6} (right) octahedral complex. The TM $s$ and $p$ states have been omitted for clarity. (b)-(c) Schematic band-edge changes as a function of strain for the paraelectric phase of (b) a \textit{trans}-ordered oxynitride with TM-N bonds only along the out-of-plane direction (c) an oxide (d) a \textit{cis}-ordered oxynitride with TM-N bonds only along the two in-plane directions.}
	\label{fig:MO}
\end{figure}

We start our analysis by determining the difference in band-gap evolution of epitaxially strained oxides and oxynitrides in the high-symmetry phase without octahedral rotations or polar distortions. To do so, we fix atoms at high symmetry positions and allow only the normal axis to relax. We will restrict our analysis to the difference in band energies between the valence-band minimum (VBM) and conduction-band maximum (CBM) which are located at the R/$\Gamma$ points of the Brillouin zone in oxides and at the $\Gamma$/$\Gamma$ points in oxynitrides, respectively.

While we confirm that for cubic oxides (i.e. SrTiO\textsubscript{3}) the band gap narrows for both compressive and tensile strain \cite{berger2011band} we find the change in band gaps of oxynitrides to depend on the anion order as shown in Fig. \ref{fig:bandgap}. For the \textit{cis} order with N-TM bonds along the in-plane direction, the band gap increases for tensile strain whereas for the \textit{trans} order with N-TM bonds along the out-of-plane direction the band gap decreases for tensile strain. This can be explained with the simple band picture already used by Berger \cite{berger2011band} for SrTiO\textsubscript{3}. By considering the simplified molecular orbital diagrams for octahedral MN\textsubscript{6} and MO\textsubscript{6} complexes shown in Fig. \ref{fig:MO}(a), we see that the highest occupied states are nonbonding N $2p$ and O $2p$ states, whereas the lowest unoccupied states are the $\pi^*$ states dominated by the TM $t_{2g}$ orbitals. We can also see that in the nitrogen complex these states are closer to the Fermi energy than in the oxygen complex. As discussed by Berger \cite{berger2011band} in a cubic oxide at zero strain the nonbonding states forming the VBM as well as the $\pi^*$ states forming the CBM are degenerate in energy (see Fig. \ref{fig:MO} c).

 For oxynitrides we expect a combination of the two molecular orbital diagrams shown in Fig. \ref{fig:MO}(a): the VBM will be dominated by N-derived nonbonding states that are closer to the Fermi energy than O-derived nonbonding states and the CBM will also contain more N than O character. Moreover the cubic symmetry is broken due to the difference between TM-O and TM-N bonds \cite{wolff2008first}, which further splits the non-bonding states forming the VBM and the $\pi^*$ states forming the CBM. In particular the degeneracy observed in oxides is already lifted at zero strain (band structures of  the \textit{trans} ordered oxynitride and the oxide showing the respective orbital contributions are shown in supplementary Figs. S6 and S7). At zero strain the anion order will result in states derived from orbitals oriented along the TM-N bonds to lie higher in energy \cite{wolff2008first}, i.e. N $2p_z$ states for the \textit{trans} order and N $2p_{x/y}$ states for the \textit{cis} order. As shown in Fig. \ref{fig:MO}(b) and (d) this results in band gaps between N $2p_z$ and TM $d_{xy}$ derived states for the \textit{trans} order as well as between N $2p_{x/y}$ and TM $d_{xz/yz}$ for the \textit{cis} order.

The strain-dependent change in energy of the individual bands edges, however, is the same as for oxides and independent of the anion order. Energies of orbitals along the out-of-plane direction remain nearly constant since the interatomic distances along that direction are little affected as strain is applied. Interatomic distances along the in-plane directions, are increased under tensile strain, which lowers the energy of orbitals along the strained directions. Conversely compressive strain will increase the orbital energies. As shown in Fig. \ref{fig:MO} (b)-(d) this places the point of degenerate VBM and CBM orbitals at compressive strain for the \textit{trans} anion order, while it is at tensile strain for the \textit{cis} anion order. As a result, we see a continuous increase in the band gap (indicated by vertical red arrows in Fig. \ref{fig:MO} (b)-(d) for the \textit{trans} order as we go from tensile to compressive strain, whereas the band gap continuously narrows for the \textit{cis} order. This model is in qualitative agreement with the DFT data shown in Fig. \ref{fig:bandgap}, which implies that the points of orbital degeneracy lie outside the strain range considered in these calculations.

\subsection{Ferroelectric distortions}\label{ferro}

\begin{figure}
	\includegraphics[width=0.9\columnwidth]{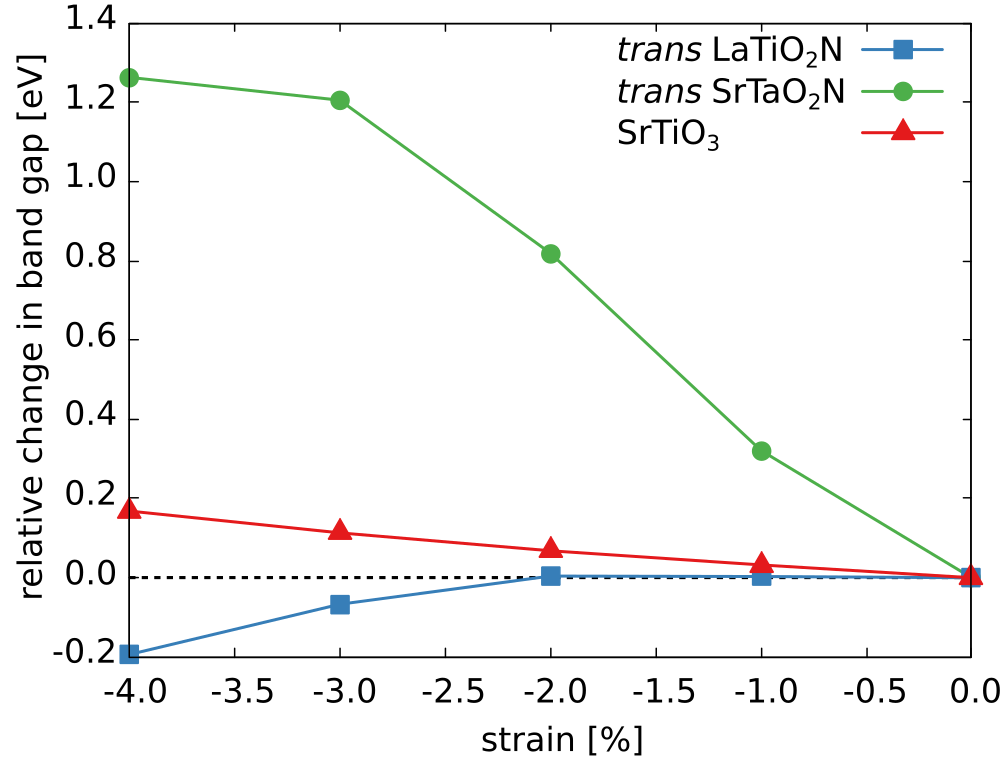}
	
	\caption{Relative change in band gap as a function of compressive strain in SrTiO\textsubscript{3}, \textit{trans} LaTiO\textsubscript{2}N and \textit{trans} SrTaO\textsubscript{2}N with ferroelectric distortions.}
	\label{fig:ferro_bandgap}
\end{figure}

We continue by studying the change in band gap with strain in the ferroelectric phase. For simplicity, we focus on the \textit{trans} anion order with N-TM bonds and a polar distortion along the out-of-plane (z) direction, which we have shown to be the energetically most stable structure under large compressive strain. The results for the \textit{cis} order are similar and we will briefly comment on them at the end of this section. In agreement with Berger \cite{berger2011band} we find a mixing of the TM \textit{d} and the anion $2p$ states as a result of the ferroelectric distortion. However, while we reproduce, as shown in Fig. \ref{fig:ferro_bandgap} and in agreement with Berger \cite{berger2011band}, a small widening of the band gap with compressive strain in SrTiO\textsubscript{3}, we observe a much larger increase in the band gap for \textit{trans}-SrTaO\textsubscript{2}N and a small decrease in the band gap for \textit{trans}-LaTiO\textsubscript{2}N.

\begin{figure}
	\includegraphics[width=0.9\columnwidth]{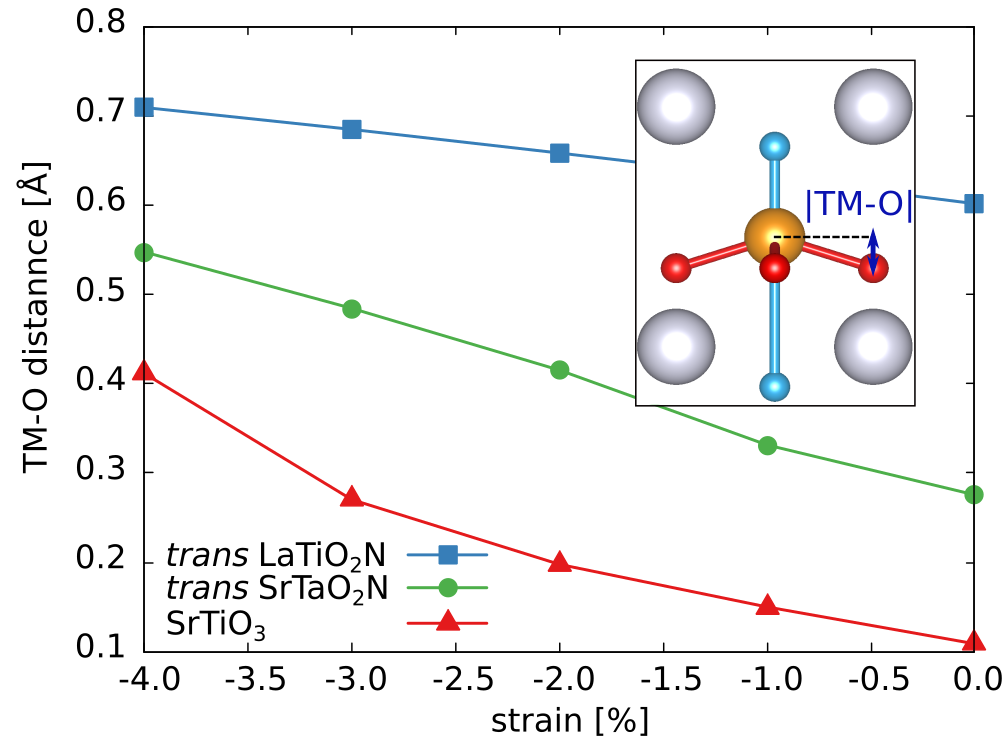}
	\caption{Distance of the transition metal from the oxygen plane (as shown in the inset) for SrTiO\textsubscript{3}, LaTiO\textsubscript{2}N and SrTaO\textsubscript{2}N as a function of compressive strain.}
	\label{fig:TMO_distance}
\end{figure}

\begin{figure}
	\includegraphics[width=0.9\columnwidth]{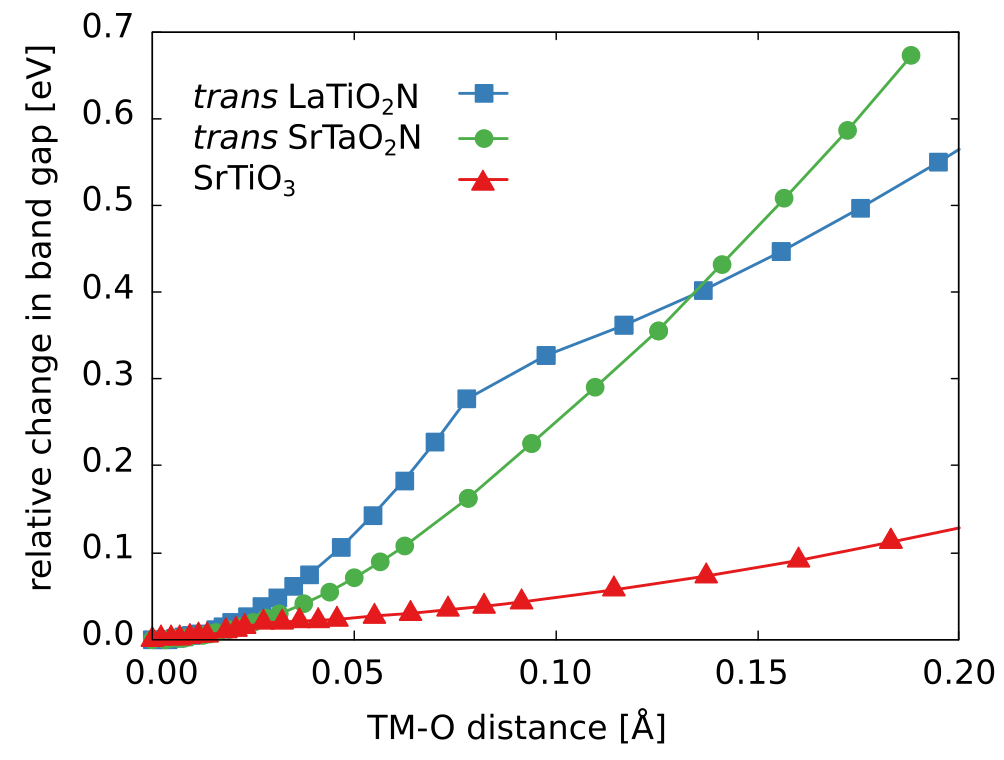}	
	\caption{Relative change in band gap as a function of the ferroelectric phonon mode amplitude (characterized by the TM-O distance, see Fig. \ref{fig:TMO_distance}) in the polar tetragonal phases of LaTiO\textsubscript{2}N, SrTaO\textsubscript{2}N and SrTiO\textsubscript{3}.}
	\label{fig:ferro_bandgap_TMO}
\end{figure}

The large difference in band-gap changes between SrTiO\textsubscript{3} and \textit{trans}-SrTaO\textsubscript{2}N with compressive strain can be explained by the absolute displacement amplitude of the ferroelectric distortion (see Fig. \ref{fig:TMO_distance}) as well as with a stronger dependence of the band gap on the distortion amplitude in oxynitrides (see Fig. \ref{fig:ferro_bandgap_TMO}) as we will discuss in the following.

We take the distance between the equatorial oxygen atoms and the transition metal along the polar distortion as a measure of the ferroelectric distortion amplitude, which is zero in a paraelectric material (see inset in Fig. \ref{fig:TMO_distance}). While the increase of the ferroelectric distortion with strain is approximately the same for both materials, the absolute value is much larger for SrTaO\textsubscript{2}N than for SrTiO\textsubscript{3} (0.27 \AA\ vs. 0.1 \AA\  at 0\% strain). We note here that unstrained SrTiO\textsubscript{3} is known to be a quantum paraelectric, in which small ferroelectric distortions are suppressed by quantum fluctuations \cite{Muller:1979wa, Zhong:1996vq}. Our GGA calculations for SrTiO\textsubscript{3} are known to overestimate the strength of the ferroelectric distortion, thus leading to a seizable displacement already at zero strain \cite{aschauer2014competition}. The difference in the magnitude of the ferroelectric distortion between SrTaO\textsubscript{2}N and SrTiO\textsubscript{3} for a certain strain can be explained with the larger ionic radius of nitrogen compared to oxygen: The larger nitrogen atom leads to an increase in the lattice parameter and hence a reduced Coulomb force, resulting in a larger amplitude of the ferroelectric distortion in oxynitrides.

Berger et al. showed that the band gap in SrTiO\textsubscript{3} increases exponentially with the amplitude of the ferroelectric distortion \cite{berger2011band}, which we confirm also for oxynitrides as shown in Fig. \ref{fig:ferro_bandgap_TMO}. We observe two exponential regimes, linked to changes in the character of the band edges. This nonlinear dependence of the band gap on the distortion amplitude implies however that materials with a larger distortion amplitude at zero strain will exhibit larger band-gap changes even if the power of the exponential change is the same. Further, we find that the power of the exponential is much larger for oxynitrides than for oxides (see Fig. \ref{fig:ferro_bandgap_TMO}). This is likely due to the enhanced TM-N covalence that results in an even faster increase of the band gap for oxynitrides than for oxides. 

We therefore, propose that the larger increase in band gap of SrTaO\textsubscript{2}N compared to SrTiO\textsubscript{3} arises from a combination of two factors. Firstly, the enhanced covalence in oxynitrides results in a steeper change of the band gap as a function of the distortion amplitude. Secondly, the exponential dependence of the band gap on the distortion amplitude combined with a larger ferroelectric distortion for SrTaO\textsubscript{2}N leads to a stronger change in band gap. 

\begin{figure}
	\includegraphics[width=0.95\columnwidth]{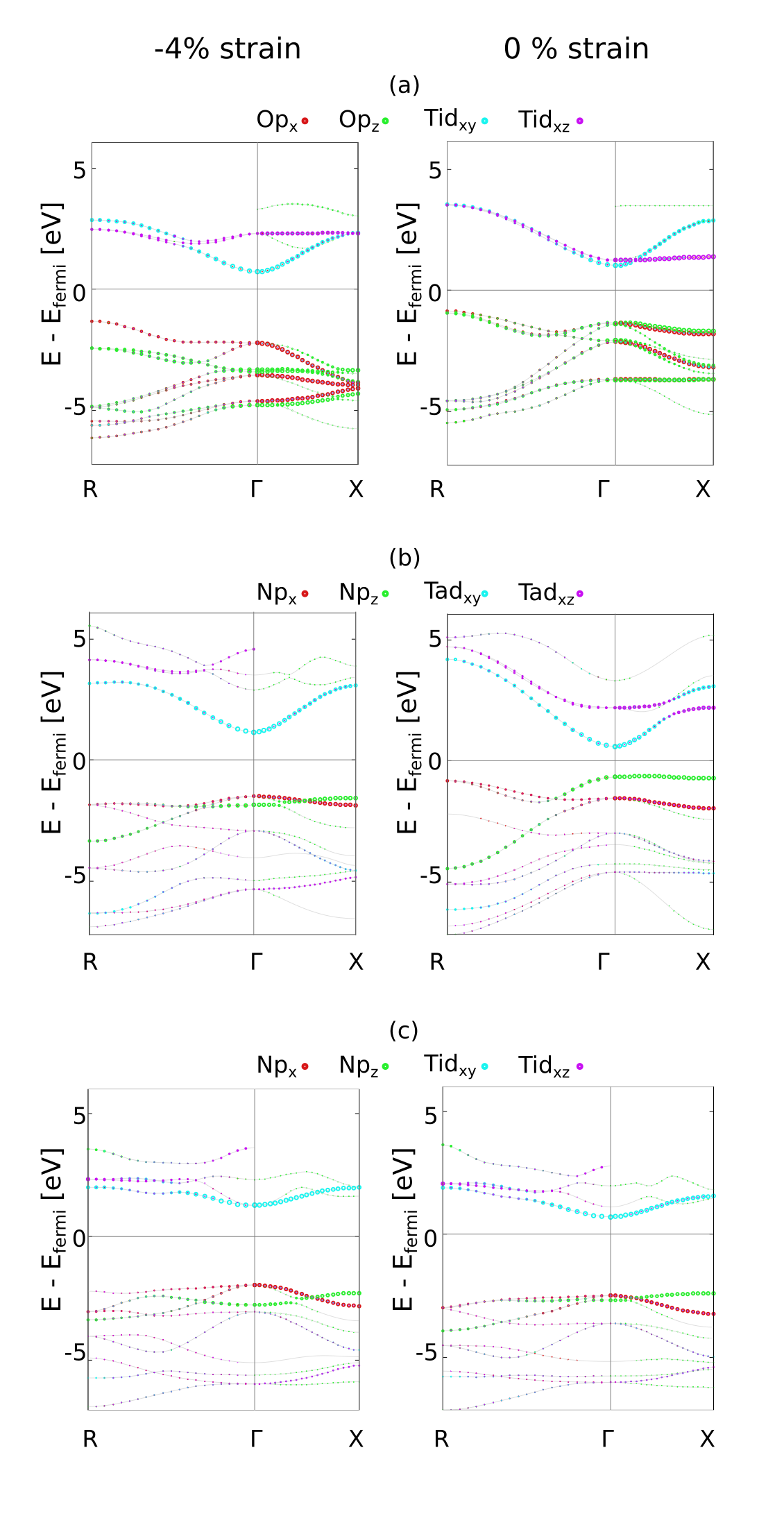}
	\caption{Fatbands for ferroelectric (a) SrTiO\textsubscript{3} , (b) SrTaO\textsubscript{2}N and (c) LaTiO\textsubscript{2}N without rotations. The labels of the points in reciprocal space refer in all three materials to the cubic system ($R=(0.5,0.5,0.5)$, $\Gamma=(0,0,0)$, $X=(0,0.5,0)$).}
	\label{fig:ferro_fatbands}
\end{figure}

For larger ferroelectric distortions we do not see an increase in the band gap anymore. For SrTaO\textsubscript{2}N we see at -4\% strain a flattening of the change in band gap and for LaTiO\textsubscript{2}N, which has larger polar distortions than SrTaO\textsubscript{2}N (see Fig. \ref{fig:TMO_distance}), we even see a decrease in the band gap over the whole strain range. We can explain this effect by a change of the orbitals that form the VBM for large ferroelectric distortions (see Fig. \ref{fig:ferro_fatbands}): The valence-band edge is dominated by the N \textit{p\textsubscript{z}} orbital for small ferroelectric distortions along the \textit{c} direction. As the ferroelectric distortion increases with increasing compressive strain, the N \textit{p\textsubscript{z}} orbital decreases in energy. At a given strain/ferroelectric distortion amplitude, the energies of the N $2p_{x/y}$ and N $2p_{y}$ orbitals will cross over and result in a change from a N $2p_{z}$-TM \textit{t\textsubscript{2g}} band gap to a N $2p_{x/y}$-TM $t_{2g}$ band gap. The band gap starts to decrease slightly at this point due to the compressive strain-induced lowering of the atomic distances in the xy-plane leading to a slight increase of the N $2p_{x/y}$ orbital as discussed in the previous section. For SrTaO\textsubscript{2}N this change in VBM orbitals occurs between 3.5 and 4\% compressive strain and results in a slower increase in band gap for this region. In LaTiO\textsubscript{2}N with a large polar distortion already at zero strain, the N $2p_{x/y}$ and N $2p_{z}$ orbitals are already nearly at the same energy without strain and therefore we observe a slight increase in the band gap for small compressive strain and a decrease of the band gap for large compressive strain.

The observed increase in band gap is hence the result of the ferroelectric distortion but a band crossover at large ferroelectric distortions can slow this increase and even lead to a decrease with increasing compressive strain and ferroelectric distortion. The analysis of the $cis$ structures with TM-N bonds in the in-plane directions leads to a similar conclusion with the difference that, here the ferroelectric distortions are along the in-plane directions and therefore increase with tensile strain \cite{vonruti2017anion}.

\subsection{Octahedral rotations}\label{oct}

So far we studied the change in band gaps for the paraelectric and ferroelectric phase without considering any rotations of the TM-O/N octahedra. However, oxynitrides were shown experimentally to have various rotation patterns \cite{clarke2002oxynitride} and we therefore turn our focus on the influence of octahedral rotations on the band gap for the paraelectric as well as for the ferroelectric phase. We first start with the simpler \textit{trans} order with nitrogen in the out-of-plane direction. The \textit{trans} structure with octahedral rotations we use here has a rotation pattern that can be described as a\textsuperscript{0}b\textsuperscript{0}c\textsuperscript{-} in Glazer notation  \cite{Glazer:1972eb}. In previous work we found this rotation pattern to be the energetically most favorable for compressive strains \cite{vonruti2017anion}. 

\begin{figure}
	\includegraphics[width=0.9\columnwidth]{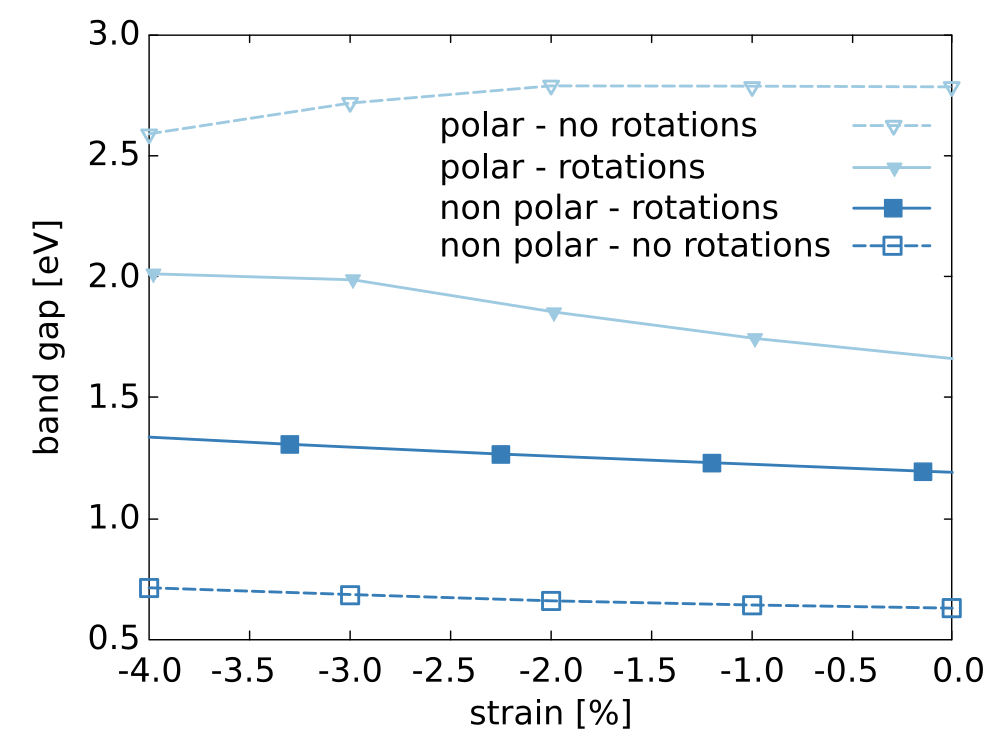}
	\caption{Band gap of \textit{trans} LaTiO\textsubscript{2}N as a function of compressive strain for different combinations of distortions.}
	\label{fig:unpolar_rotations}
\end{figure}

\begin{figure}
	\includegraphics[width=0.9\columnwidth]{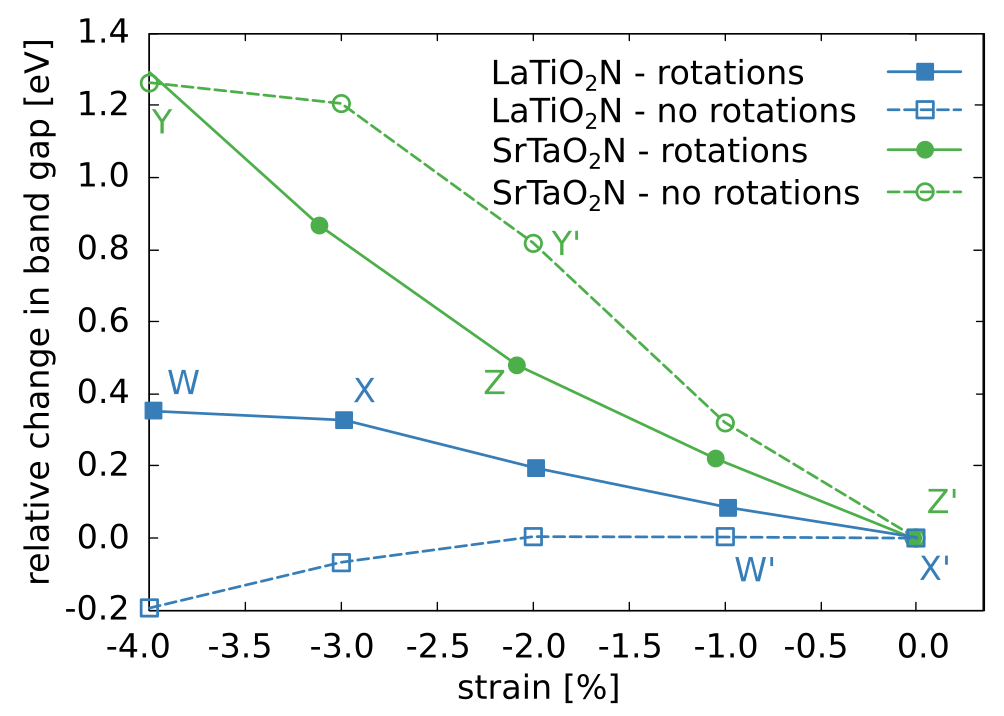}
	\caption{Relative change in band gap as a function of compressive strain for ferroelectric \textit{trans} SrTaO\textsubscript{2}N and LaTiO\textsubscript{2}N with and without octahedral rotations.}
	\label{fig:polar_rotations}
\end{figure}

For paraelectric LaTiO\textsubscript{2}N we observe an increase of the absolute value of the band gap by introducing rotations (see Fig. \ref{fig:unpolar_rotations}). This is in agreement with Berger \cite{berger2011band} who found that similar to ferroelectric distortions, octahedral rotations lead to a mixing of TM and O states in oxides resulting in a widening of the band gap. On the other hand, when we compare the absolute value of the band gap of the ferroelectric phase with and without octahedral rotations we find that octahedral rotations decrease the band gap. Ferroelectric distortions and octahedral distortions were shown to often compete \cite{Benedek:2013jl, aschauer2014competition} and octahedral rotation thus reduce the amplitude of a polar distortion. Given that the polar distortion has a larger effect on the band gap than octahedral rotations, the combined effect of both distortions is still an increase, however of smaller magnitude because the amplitude of the polar distortion is reduced by the octahedral rotations. This can be also seen by looking at the relative change of the band gaps for the ferroelectric phase with and without rotations (see Fig. \ref{fig:polar_rotations}). We have seen in the previous section that for small ferroelectric distortions the band gap increases with increasing amplitude of the distortions and starts to decrease when the N $p_z$ orbital reaches the energy of the N $p_x$ and N $p_y$ orbital. The partial suppression of the polar distortion by octahedral rotations shifts this crossover to larger strains as can be seen by the similar evolution of the SrTaO\textsubscript{2}N band gap between point Y' and Z' without rotations compared to Y and Z with rotations. The same is true for LaTiO\textsubscript{2}N where the downturn in band gap is shifted away from zero strain to larger compressive strains (compare segment W'-X' without rotations to segment W-X with rotations).


For the \textit{cis} anion order the investigation is slightly problematic since it is often not possible to let the atomic positions relax while maintaining a paraelectric structure with rotations. However, from our results we do not see any reason why the \textit{cis} structure should behave differently from the \textit{trans} structure.

\section{Conclusions}

We find that for the paraelectric phase the magnitude of the change in band gap with strain is approximately the same for oxides and oxynitrides. However, while for oxides the band gap decreases slightly under both compressive and tensile strain, for oxynitrides the sign of the change depends on the anion order and the resulting alignment of the TM-N bonds with the strain axis. Further, we find that in oxynitrides, ferroelectricity increases the band gap much more than in oxides due to two reasons: First, oxynitrides show a larger polar displacement than oxides because the larger atomic radius of nitrogen leads to expanded lattice parameters and hence stronger polar instabilities compared to oxides. Second, there is a stronger dependence of the band gap on the amplitude of the ferroelectric distortions in oxynitrides due to larger energetic overlap of the N $2p$ and TM $d$ orbitals. Octahedral rotations play a smaller role that manifests mostly in their partial suppression of the polar distortion and therefore a reduction of the aforementioned effects.

Using this knowledge we can now explain the evolution of the band gap in the thermodynamically stable phases of SrTaO\textsubscript{2}N and LaTiO\textsubscript{2}N (see Fig. \ref{fig:bandgap_real}). SrTaO\textsubscript{2}N due to its 5d orbitals has stronger covalent TM-N bonds compared to LaTiO\textsubscript{2}N (3d orbitals), which will render all thermodynamical stable phases ferroelectric \cite{kubo2017mgtao2n,kubo2017anion, ziani2017photophysical}. This leads to the observation of a large increase in band gap for all described phases. Contrary, LaTiO\textsubscript{2}N has in an intermediate strain range phases that are not ferroelectric ($cis_{bc}^{ac}$ and $cis_{ac}^{ac}$) and which show only a slight change in band gap while the others are ferroelectric and show a large change in band gap. 

Since ferroelectric distortions occur in SrTaO\textsubscript{2}N for all strain ranges (including unstrained bulk) and in LaTiO\textsubscript{2}N \cite{vonruti2017anion} for both compressive and tensile strain, the resulting increase in band gap could adversely affect the performance of oxynitride photocatalysts. The suppression of ferroelectric distortions, for example by increasing temperature, might therefore significantly decrease the band gap and lead to an increased efficiency even for small strains or bulk samples in the case of SrTaO\textsubscript{2}N and other tantalum based oxynitrides \cite{kubo2017mgtao2n,kubo2017anion}. Also while high compressive strain could enhance the carrier separation in oxynitrides \cite{oka2014possible,oka2017strain,vonruti2017anion}, it will at the same time lead to an increased band gap with detrimental effects on the photocatalysis.

\section{Acknowledgements}
This research was funded by the SNF Professorship Grant PP00P2\_157615. Calculations were performed on UBELIX (http://www.id.unibe.ch/hpc), the HPC cluster at the University of Bern.

\bibliography{bib2}

\end{document}